# Blue-phase-polymer-templated nematic with sub-millisecond broad-temperature range electro-optic switching

Jie Xiang and Oleg D. Lavrentovich[*]

*Liquid Crystal Institute and Chemical Physics Interdisciplinary Program,
Kent State University, Kent, USA*

We report on fast electro-optic switching (response time 0.1 ms) of a blue-phase-polymer templated nematic with a broad-temperature range of thermodynamic stability and hysteresis-free performance. The nematic fills a polymer template that imposes a periodic structure with cubic symmetry and submicron period. In the field-free state, the nematic in polymer template is optically isotropic. An applied electric field causes non-zero optical retardance. The approach thus combines beneficial structural and optical features of the blue phase (cubic structure with submicron periodicity) and superior thermodynamic stability and electro-optic switching ability of the nematic filler.

Blue phases (BPs) of liquid crystals (LCs) represent an example of a frustrated soft matter system[1,2]. They are formed by chiral molecules that tend to arrange locally into structures with two axes of twist (so-called double twist). Although locally the double-twist is preferable than the single-twist structure (typical of standard cholesteric LCs), it cannot extend itself to fill the entire volume and needs to be stabilized by a lattice of topological defects-disclinations[1,2]. At the cores of disclinations, orientational order is reduced, so that the material can be considered as partially melted. This is why the BPs are typically observed only within a close proximity (one-two degrees) of the isotropic phase. Depending on arrangements of ordered and disordered regions, one distinguishes three classes of the BPs: BPI (body-centered cubic structure), BPII (simple cubic structure) and BPIII (amorphous lattice)[1,2]. In absence of an electric field, the BPs are optically isotropic, i.e., they show no birefringence. An applied electric field $E$ lifts the symmetry and causes birefringence $\Delta n_E = \lambda K E^2$ that is described as a Kerr effect with the Kerr constant $K$, generally on the order of $(10^{-10}$-$10^{-9})$ m/V$^2$; $\lambda$ is the wavelength of probing light[3,4].

There are several advantages of using BPs in electro-optic applications, as compared to the traditional nematic LCs, such as absence of alignment layers and fast (sub-millisecond) optical response to the switched electric field[3]. However, practical applications are hindered by intrinsic problems, the most serious of which is a narrow temperature range of BPs.

Two approaches have been demonstrated to extend the BPs temperature range[5-7]. The first one is a polymer-stabilized blue phase (PSBP)[5,7]. One of its drawbacks is hysteresis of electrooptic response[8]. The second approach is to search for new LC molecules, such as bimesogens[6], that yield beneficial material parameters such as enhanced flexoelectricity[9] and reduced bend elastic constant[10,11]. However, the bimesogenic materials have a relatively low dielectric anisotropy and high viscosity[12], which leads to a longer response time and higher driving voltages[6]. One might hope to formulate a suitable PSBP or BP composition by using a broad-temperature range non-chiral nematic and doping it with a chiral additive, but the latter often reduces the temperature range of the resulting mixture and makes the electrooptic performance very temperature-sensitive[13]. The status of current research makes it clear that the material properties that are beneficial for the temperature stability of BPs are not necessarily beneficial for electro-optic performance. In this work, we propose an electro-optically switchable BP-polymer-templated nematic (BPTN) in which the specially formulated BP mixture is used only to create a polymer template, but is not used for the electro-optical switching; the latter function is performed by a new filler in the polymer template, such as a broad-temperature range nematic.

---

[*] Electronic mail: olavrent@kent.edu

The general idea of mutual LC-polymer templating (transferring LC structure to polymers through photo-polymerization and imposing structural changes in the LCs by the polymer network) has been already explored, see, e.g., Refs.[14,15]. Experimentally, attraction of a polymerizing material to the disclination core has been demonstrated for disclinations in a twisted nematic[16]. Higashiguchi et al.[17] observed spatially periodic polymer networks formed in BPs. Guo et al.[18] described a washout refill method to create hyper-reflective chiral nematics, in which the unpolymerized components are washed out and the residual polymer network is refilled with another LC. Castles et al.[19] used bimesogenic LCs to prepare a BP-templated polymer network and then refilled it with various nematics, to achieve unprecedented temperature stability (from -125 to 125$^{0}$C) and mirrorless lasing. The possibility of electro-optic switching was also demonstrated, although the merit parameters such as switching times and switching amplitude of optical retardance were not characterized.

In this work, we apply the templating approach to create a BPTN, a fast switching broad-temperature range electro-optical material. First, the BP was formed in a mixture of commercially available materials, a nematic MLC2048, chiral dopant S811 (both purchased from EM Industries), reactive monomers RM257 (BDH, Ltd) and TMPTA (Aldrich), and photoiniator IRG651 (Aldrich) with weight percentages 51wt%, 36.1wt%, 7.3wt%, 5wt%, and 0.6wt%, respectively. The mixture was injected into the glass cell of thickness 3.8μm in its isotropic phase. The material was then cooled (0.2$^{o}$C/min) to 24$^{o}$C at which it shows a supercooled BPI phase. The mixture was irradiated with UV light (wavelength 365nm, intensity 1mW/cm$^2$) for 3 hours to induce photopolymerization. The texture of the resulting PSBP under crossed polarizer is shown in Fig.1(b).

Second, the PSBP was placed in hexane for 20 hours, in order to remove the un-polymerized components, Fig.1(c). The residual hexane was evaporated by drying at room temperature.

To visualize the polymer network, we used a scanning electron microscope (SEM) Hitachi S-2600N, Fig.1(d). One of the plates of the cell was removed and a thin (~20 nm) layer of gold was deposited onto the polymer structure to enhance the contrast. The SEM texture clearly shows a periodic structure with a typical size of pores around 200 nm, as shaped by the original periodic cubic structure of BPI, Fig.1 (e,f).

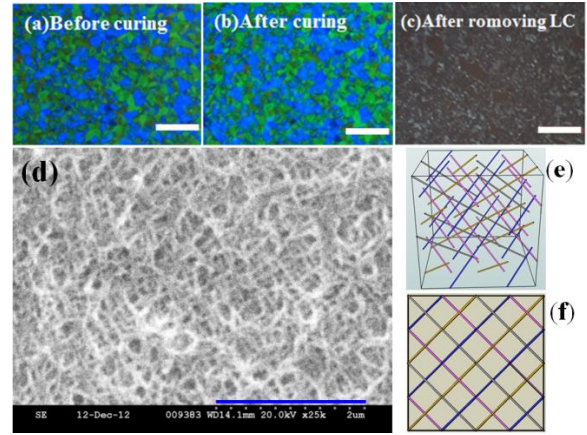

FIG.1. Polarizing optical microscope textures of BP mixture (a) before and (b) after photo-polymerization, (c) after removing the LC; (d) SEM texture of BP-shaped polymer network; (e) 3D arrangement of disclinations in BPI; (f) 2D projection of the disclination network in (e). The scale bar is 100 μm in (a)-(c), and 2μm in (d).

Finally, the template was refilled with a broad-temperature nematic mixture designed to perform electro-optic switching. The polymer template transfers the BP structural organization onto the new filler, resulting in a BPTN. Depending on the particular application, the filler can be designed to have a high dielectric anisotropy (to reduce the operational voltage), high birefringence (to increase the switchable phase retardance), extended temperature range of stability, etc. As an example, we use a standard nematic mixture E7 (EM Industries). The nematic phase of E7 is stable between -30$^{o}$C and 58$^{o}$C[20]. The material is nonchiral and does not form the BPs by itself. However, once E7 is inside the polymer template, it shows textures very similar to those of the BPs, in a very broad temperature range, from -30$^{o}$C (the lowest temperature we used in the experiments) to 60$^{o}$C, Fig.2, which is much wider than the range of the original BP (42$^{o}$C-22$^{o}$C on cooling).

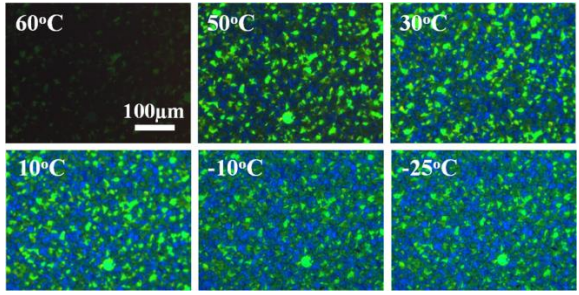

FIG.2. Polarizing optical microscope textures of BPTN E7 at different temperatures.

To study the electro-optical performance, we used in-plane-switching (IPS) cells. The IPS cells of thickness of 3.8 µm were formed by two parallel glass substrates, one with patterned indium tin oxide (ITO) electrodes (stripe electrodes of 10 µm width and 10 µm spacing). The textural response to the in-plane AC electric field (rectangular pulses of frequency $f = 10$ KHz) is shown in Fig. 3.

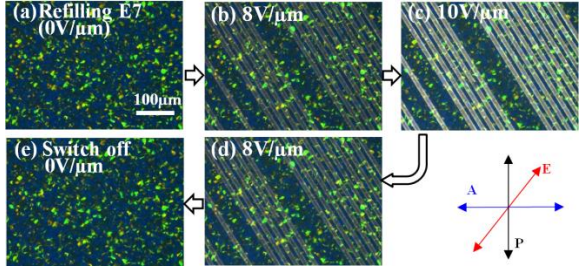

FIG.3. Polarizing microscope textures of BPTN E7 cell under AC electric field ($f = 10$ KHz) directed at 45 degree with respect to the two crossed polarizes labeled A and P. Room temperature.

Figure 4(a) shows the optical phase retardance of BPTN E7 cells as a function of the applied AC electric field ($f = 10$ KHz) for three consecutive driving cycles. The retardance is measured by LC-PolScope (Cambridge Research and Instrumentation)[21,22] at wavelength $\lambda = 546$ nm. The electro-optic switching is practically hysteresis-free, which is a significant advantage as compared to the regular PSBP switching[8]. The hysteresis, defined as the voltage difference at half-maximum phase retardance change between forward and backward directions, is $\Delta V = 0.77$ V. As compared to the maximum applied voltage $V_P$, it is very small, $\Delta V / V_P = 0.64\%$. Figure 4(b) shows the temperature dependence of Kerr constant $K$. It is of the same order as reported for the best PSBPs[7,23], with the advantage that the order of magnitude of $K$ is preserved even at -10°C. As clear from the dependence[24] $K \propto \Delta n \cdot \Delta \varepsilon$, Kerr constant can be further increased by using a nematic filler with a large dielectric anisotropy $\Delta \varepsilon$ and field-free birefringence $\Delta n$ [25]; for E7, these parameters are rather average, $\Delta \varepsilon = 13.8$ and $\Delta n \approx 0.2$; much higher values are currently available, e.g., $\Delta \varepsilon$ in excess of 100 [26].

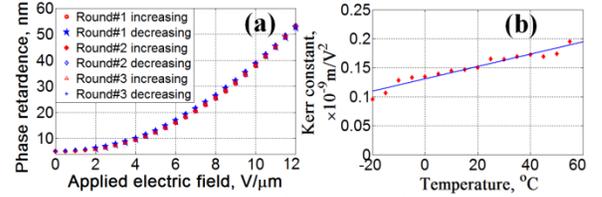

FIG.4. Electro-optical performance of 3.8µm IPS BP-templated E7 cell. (a) Phase retardation vs the applied AC field; $f = 10$ KHz; three cycles of driving at 25°C. (b) Temperature dependence of Kerr constant. The line is a guide to an eye.

We determined the characteristic times of the electro-optic response by recording the change of light transmittance through the cells and a pair of crossed polarizers and determining the levels of 10% and 90% of the maximum transmittance. Both switch-on $\tau_{on}$ and switch-off $\tau_{off}$ times are below 0.1 ms when the cells are driven by AC voltage (rectangular pulses) of amplitude 100 V and frequencies 1 KHz and 125 Hz, in the temperature range (30-50)°C, Fig.5(c) and Table. Most importantly, these times remain shorter than 1 ms when the temperature is as low as (-20)°C, Fig.5(c). The response times are close to the record numbers achieved with PSBPs for oblique light propagation and vertical electric addressing (e.g., $\tau_{off}$ =39 µs at room temperature[23]), and are much smaller than the response times in standard BPs driven by IPS (~0.5 ms[7,27]).

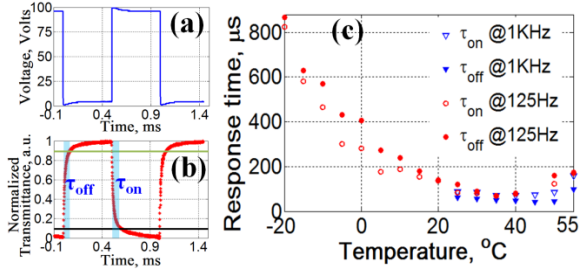

FIG.5. Electro-optic response of BP templated E7: (a) applied voltage profile with frequency 1 KHz and (b) corresponding light transmittance at 25°C; (c) Temperature dependence of the on- and off- response times for an applied voltage 100 V at different frequencies.

To illustrate the advantages of our approach in terms of thermal stability and electro-optical switching, we compare the performance of BPTN with that of the original un-polymerized BP and PSBP (with the initial BP material remaining as a filler after the polymer network has been formed), Table I. The response times in all three cases were measured in the same manner, by applying a rectangular-wave voltage with magnitude 100V and frequency 1 KHz in the IPS cells. As compared to the pure BP and PSBP, the BPTN approach offers the widest BP temperature range, the largest Kerr constant, and also features a fast sub-millisecond electro-optic switching at temperatures at which BP and PSBP are not switchable at all.

TABLE I: Comparison of pure BP, polymer-stabilized BP and BP-polymer-templated nematic approaches.

| | BP range (°C) | Kerr constant ($\times 10^{-9} m/V^2$) | Response time Switch-on | Response time Switch-off |
|---|---|---|---|---|
| Pure BP | 54 -61 °C | 0.14 (at 56 °C) | 61μs (at 56 °C) | 82 μs (at 56 °C) |
| PSBP | <-30 °C - 56 °C | 0.014 (at 25 °C) | 154μs (at 25 °C) | 180μs (at 25 °C) |
| BPTN filled E7 | <-30 °C - 60 °C | 0.16 (at 25 °C) | 87μs (at 25 °C) | 61μs (at 25 °C) |

In conclusion, we demonstrated an electro-optically switchable blue-phase-polymer-templated nematic (BPTN) that combines the advantages pertinent to the BP phase such as fast switching for both field on and field off driving, absence of surface alignment layers, zero birefringence/retardance ground state, with the advantages of the nematic LCs, such as broad temperature range of existence, absence of hysteresis, and fast switching of optical retardance.

**Acknowledgements.** This work was supported by DOE Grant No. DE-FG0206ER 46331 (electro-optic studies) and NSF grant DMR 1121288 (template preparation). We thank Nickolas Abbott, Volodymyr Borshch, Liang-Chy Chien, and Deng-ke Yang for fruitful discussions and Liu Qiu for SEM imaging.

**References:**

[1] P. G. de Gennes and J. Prost, *The physics of liquid crystals*, 2nd ed (Oxford University Press, Oxford, New York, 1993), p.328.
[2] M. Kleman and O. D. Lavrentovich, *Soft matter physics: an introduction* (Springer, New York, 2003), p.407.
[3] H. S. Kitzerow, Ferroelectrics **395**, 66 (2010).
[4] H. S. Chen, S. Y. Ni, and Y. H. Lin, Appl. Phys. Lett. **101**, 093501 (2012).
[5] H. Kikuchi, M. Yokota, Y. Hisakado, H. Yang, and T. Kajiyama, Nature Mater. **1**, 64 (2002).
[6] H. J. Coles and M. N. Pivnenko, Nature **436**, 997 (2005).
[7] H. Choi, H. Higuchi, Y. Ogawa, and H. Kikuchi, Appl. Phys. Lett. **101**, 131904 (2012).
[8] K. M. Chen, S. Gauza, H. Q. Xianyu, and S. T. Wu, J Display Technol. **6**, 318 (2010).
[9] F. Castles, S. M. Morris, E. M. Terentjev, and H. J. Coles, Phys. Rev. Lett. **104**, 157801 (2010).
[10] S. T. Hur, M. J. Gim, H. J. Yoo, S. W. Choi, and H. Takezoe, Soft Matter **7**, 8800 (2011).
[11] J. I. Fukuda, Phys. Rev. E **85**, 029903 (2012).
[12] S. M. Morris, M. J. Clarke, A. E. Blatch, and H. J. Coles, Phys. Rev. E **75**, 041701 (2007).
[13] L. Rao, J. Yan, S. T. Wu, S. Yamamoto, and Y. Haseba, Appl. Phys. Lett. **98**, 081109 (2011).
[14] Y. K. Fung, D. K. Yang, S. Ying, L. C. Chien, S. Žumer, and J. W. Doane, Liq. Cryst. **19**, 797 (1995).
[15] Y. C. Yang and D. K. Yang, Appl. Phys. Lett. **98**, 023502 (2011).
[16] D. Voloschenko, O. P. Pishnyak, S. V. Shiyanovskii, and O. D. Lavrentovich, Phys. Rev. E **65**, 060701 (2002).
[17] K. Higashiguchi, K. Yasui, M. Ozawa, K. Odoi, and H. Kikuchi, Polymer Journal **44**, 632 (2012).
[18] J. B. Guo, H. Cao, J. Wei, D. W. Zhang, F. Liu, G. H. Pan, D. Y. Zhao, W. L. He, and H. Yang, Appl. Phys. Lett. **93**, 201901 (2008).
[19] F. Castles, F. V. Day, S. M. Morris, D. H. Ko, D. J. Gardiner, M. M. Qasim, S. Nosheen, P. J. W. Hands, S. S. Choi, R. H. Friend, and H. J. Coles, Nature Mater. **11**, 599 (2012).

[20]P. Sathyanarayana, M. C. Varia, A. K. Prajapati, B. Kundu, V. S. S. Sastry, and S. Dhara, Phys. Rev. E **82**, 050701 (2010).

[21]R. Oldenbourg and G. Mei, J Microsc-Oxford **180**, 140 (1995).

[22]M. Shribak and R. Oldenbourg, Appl. Optics **42**, 3009 (2003).

[23]Y. Chen, J. Yan, J. Sun, S. T. Wu, X. Liang, S. H. Liu, P. J. Hsieh, K. L. Cheng, and J. W. Shiu, Appl. Phys. Lett. **99**, 201105 (2011).

[24]P. R. Gerber, Mol. Cryst. Liq. Cryst. **116**, 197 (1985).

[25]J. Yan and S. T. Wu, Opt. Mater. Express **1**, 1527 (2011).

[26]B. Ringstrand and P. Kaszynski, J. Mater. Chem. **21**, 90 (2011).

[27]K. M. Chen, S. Gauza, H. Q. Xianyu, and S. T. Wu, J Display Technol. **6**, 49 (2010).